\documentstyle[twocolumn,epsfig,aps]{revtex}
\begin{document}
\setlength{\topmargin}{-0.05in}
 \draft
\title{Creating maximally entangled atomic states in a
Bose-Einstein condensate}
\author{L. You}
\address{School of Physics, Georgia Institute of Technology, Atlanta GA
30332, USA}
\address{Institute of Theoretical Physics, The Chinese
Academy of Sciences, Beijing 100080, People's Republic of China}
\date{\today}
\maketitle
\begin{abstract}
We propose a protocol to create maximally entangled pairs,
triplets, quartiles, and other clusters of Bose condensed
atoms starting from a condensate in the
Mott insulator state. The essential element is to drive
single atom Raman transitions using laser pulses.
Our scheme is simple, efficient, and can be readily applied
to the recent experimental system as reported by
Greiner {\it et al.} [ Nature {\bf 413}, 44 (2002)].
\end{abstract}


\narrowtext


The physics of quantum degenerate atomic gases
continues in its rapid pace of development, and
remains one of the most active research areas
in recent years \cite{Edwards}.
Increasingly, theoretical and experimental attentions
are directed towards the underlying quantum correlation
properties of the condensed atoms.
It seems likely that such quantum states of matter
might prove to be a fertile ground for exploring quantum
information science applications.

Recently, a quantum phase transition was observed
in a system of Bose condensed atoms immersed in
a periodic array of optical potentials \cite{hansch}.
As expected, when expressed in the simple
Bose-Hubbard form \cite{zoller}, the ground state of
such a system
is controlled by essentially two parameters:
1) the on-site atom-atom interaction $u$ for
atoms in the same spatial mode
of each individual optical well;
and 2) the nearest neighboring
well (single) atom tunnelling rate $J$ (taken as positive).
When $J\gg |u|$, the condensate ground state is in
the usual superfluid (delocalized single atom) state.
On the other hand,
a Mott insulator state arrives in the opposite
limit $|u|\gg J$.
In a Mott state, atoms are localized inside individual
wells. The condensate ground state takes the form of
a product of Fock states with an integer number of atoms on each site.
The transition from superfluid to Mott insulator is
predicted to occur at
$|u|/ J\ge {\sf z}\times 2.6$
with ${\sf z}$ the number of nearest neighbors
in the periodic well lattice \cite{zoller,sa}.

The experimental system that yielded the first
clear demonstration of the superfluid/Mott-insulator
transition enables individual tuning of
the values for both $J$ and $u$ \cite{hansch}.
In the experiment, the average
occupations per well was around 1-3 atoms,
which could potentially form elementary building
blocks for atomic qubit based quantum computing
designs \cite{zoller}.

In this paper, we
propose to create massive maximum entangled pairs,
triplets, quartiles, and other clusters of Bose condensed
atoms in a Mott insulator state.
The resulting entanglement, with respect
to electronic internal states, is stable and
long lived. In the experiment \cite{hansch}
$^{87}$Rb atoms in the magnetic trapping state
$|a\rangle=|F=2,M_F=-2\rangle$ were used.
Other internal states can be trapped in the
same experimental setup.
In the simple model to be presented below,
a second internal state $|b\rangle$
that can be coupled to $|a\rangle$
through atomic Raman transitions is assumed \cite{hansch}
(as see earlier JILA experiments with $^{87}$Rb
states $|F=2,M_F=-1\rangle$ and $|F=1,M_F=1\rangle$ \cite{jila}).

In a Mott state, the system dynamics
is rather simple as there exists a fixed (small)
number of atoms within each well. If we
use the second quantized operators $a (a^\dag)$ and $b(b^\dag)$
for atoms in the two internal states, the
effective Hamiltonian for each well can be expressed as \cite{sorensen}
\begin{eqnarray}
H=uJ_z^2+\Omega J_y.
\label{h1}
\end{eqnarray}
The second term denotes the single atom Raman coupling
due to external laser fields with a (real)
effective Rabi frequency $\Omega(t)$ \cite{law}.
The angular momentum operators are the
Schwinger representation in terms of the two boson modes
\begin{eqnarray}
J_x &&= {1\over 2}(b^\dag a+a^\dag b),\nonumber\\
J_y &&= -{i\over 2}(b^\dag a-a^\dag b),\nonumber\\
J_z &&= {1\over 2}(b^\dag b-a^\dag a).
\end{eqnarray}
In the context of SU(2) coherent states of an atomic ensemble,
these operators have been
used extensively for discussing spin squeezing
and other properties of multi-atom nonclassical states
\cite{kitagawa,wineland,yale,kris}.
In particular, as was studied by Molmer
and Sorensen \cite{mol}, an interaction of the type
$uJ_x^2$ generates
a maximum entangled N-GHZ state \cite{ghsz}
starting from all atoms
in state $|a\rangle$ or $|b\rangle$.
This has led to the recent creation
of a 4-ion maximum entangled state \cite{monroe}.

Before we discuss our proposal, we summarize the dynamic
generation of a maximum entangled state from the $uJ_x^2$
interaction. For simplicity, we assume $N$ is even.
A maximum entangled N-GHZ state can be written as \cite{mol}
\begin{eqnarray}
|{\rm GHZ}\rangle_N =&&{1\over \sqrt 2}\left(
e^{i\phi_b}{b^{\dag N}\over \sqrt{N!}}
+e^{i\phi_a}{a^{\dag N}\over \sqrt{N!}}\right)|0\rangle\nonumber\\
=&&{1 \over 2^{{N+1}\over 2}\sqrt{N!} }
\sum_{m=-{N\over 2}}^{{N\over 2}}C_N^{{N\over 2}+m}
d^{\dag {N\over 2}+m}c^{\dag {N\over 2}-m}\nonumber\\
&&[e^{i\phi_b}+e^{i\phi_a}(-1)^{{N\over 2}-m}]|0\rangle,
\label{ghz}
\end{eqnarray}
where new bosonic operators $d/c=(b\pm a)/\sqrt{2}$
were introduced along with its inverse
$b/a=(d\pm c)/\sqrt{2}$.
$C_N^M$ is the binomial coefficient.
Starting from all atoms in state $|a\rangle$,
i.e. with $|\psi(0)\rangle={a^{\dag N}|0\rangle/\sqrt{N!}}$.
The state at time $t$ due to a
$uJ_x^2$ interaction alone is
\begin{eqnarray}
|\psi(t)\rangle =&&{1\over 2^{N\over 2}\sqrt{N!}}
\sum_{m=-{N\over 2}}^{N\over 2}C_N^{{N\over 2}+m}
d^{\dag {N\over 2}+m}c^{\dag {N\over 2}-m}\nonumber\\
&&e^{-iut m^2}(-1)^{{N\over 2}-m}|0\rangle,
\end{eqnarray}
where use has been made of $J_x=(d^\dag d-c^\dag c)/2$.
To within an overall phase factor
$|\psi(\tau)\rangle\equiv |{\rm GHZ}\rangle_N$
at $u\tau=(2k+1)\pi/2$ with the shortest time being $\tau=\pi/(2|u|)$.
Similarly, starting from state $b^{\dag N}|0\rangle$ will
also arrive at a N-GHZ when $u\tau=(2k+1)\pi/2$ \cite{note}.

How could interaction (\ref{h1}) be turned into the
required $J_x^2$ form? Our key observation is
that the single atom Raman coupling $\Omega J_y$
generates nothing but a rotation along the y-axis.
Therefore, we can effectively rotate the
$J_z^2$ term into a $J_x^2$ term. A similar suggestion was made
recently by Jaksch {\it et. al.} \cite{jaksch}
in order to tune the overall condensate interaction
strength to zero (or SU(2) symmetric).

We therefore suggest operating in a three step protocol
in the limit when $|\Omega|\gg N|u|$:

1) Apply a $\pi/2$ pulse $\theta(\tau')=\int_0^{\tau'}\Omega(t)dt=\pi/2$
(of spin 1/2).
During this stage the nonlinear interaction
can be neglected (because $|\Omega|\gg N|u|$).

2) Wait for a time $|u|\tau=\pi/2$.

3) Complete the process by applying a
$-\pi/2$ pulse with $\theta(\tau')=-\pi/2$
[e.g. by arranging for $\Omega\to-\Omega$
or by waiting for a $3\pi/2$ pulse as in 1) ].

These three steps generate the following
effective evolution,
\begin{eqnarray}
U(2\tau'+\tau) &&\approx e^{i{\pi\over 2} J_y}
e^{-i{\pi\over 2} J_z^2}e^{-i{\pi\over 2} J_y}=e^{-i{\pi\over 2} J_x^2},
\end{eqnarray}
i.e. $J_z^2$ is rotated by $\pi/2$ into
$J_x^2$. From a wide range of numerical simulations,
we find that N-GHZ states with
extremely high fidelities are realized when
$|\Omega|/|u|\ge 50$ for (up to 4 atoms).

While the above scheme works well,
it is inherently rather slow.
In a two component condensate
as assumed, we denote the 3 relevant scattering
lengths as $a_{aa}$, $a_{ab}$, and $a_{bb}$,
and assume that motional ground state
to be $\psi_{000}(\vec r)=
\exp[-r^2/(4 a_h^2)]/(\sqrt{2\pi}\, a_h)^{3/2}$
of a spherically
symmetric harmonic trap $V(\vec r)=M\omega_t^2 r^2/2$,
we find
\begin{eqnarray}
u=(a_{aa}+a_{bb}-2a_{ab}){2\pi\hbar^2\over M}{1\over (2\sqrt{\pi} a_h)^3},
\end{eqnarray}
with $a_h=\sqrt{\hbar/2M\omega_t}$ the ground state size.
For $^{87}$Rb, $u$ is very small as
$a_{aa}\sim a_{ab}\sim a_{bb}$.
When $\omega_t\sim (2\pi) 30$ (kHz) as
realized in \cite{hansch}, $|u|\sim (2\pi) 20$ (Hz)
if $(a_{aa}+a_{bb}-2a_{ab})$ is of the order of
one Angstrom ($\AA$). It takes approximately
$10$ (ms) to realize a GHZ state, i.e. in a time
significantly shorter than the lifetimes from both
the two-body dipolar [$>6$(s)] and the
three-body inelastic collision [$>200$(ms)] losses
with less than five atoms in each well \cite{carl}.

Another serious experimental concern is
that collisions can populate Zeemen states
other than $|a\rangle$ or $|b\rangle$.
For most systems,
this depopulation also occurs on the time scale
of $\sim 1/|u|$.
It is therefore important to include the full manifold
of atomic internal states.
To this end, we consider a
spinor-1 condensate of $^{87}$Rb atoms in its
ground state $F=1$ manifold as realized in the
first all optical condensate \cite{chapman}.
If $a_{M_F}$ denotes the bosonic annihilation
operator of state $|F=1,M_F=+,0,-\rangle$,
the ground state Hamiltonian within each well becomes
\begin{eqnarray}
H'&&=u(L^2-2N) \nonumber\\
&&=u(a_+^\dag a_+^\dag a_+a_++a_-^\dag a_-^\dag a_-a_-\nonumber\\
&&+2a_+^\dag a_0^\dag a_+a_0+2a_-^\dag a_0^\dag a_-a_0
-2a_+^\dag a_-^\dag a_+a_-\nonumber\\
&&+2a_0^\dag a_0^\dag a_+a_-+2a_+^\dag a_-^\dag a_0a_0),
\end{eqnarray}
with angular momentum type operators \cite{pu,ueda,ho}
\begin{eqnarray}
L_+ &&=\sqrt{2}(a_+^\dag a_0+a_0^\dag a_-), \hskip 12pt L_-=L_+^\dag, \nonumber\\
L_z &&=a_+^\dag a_+-a_-^\dag a_-,
\end{eqnarray}
and number of atoms in
the well $N=a_+^\dag a_++a_0^\dag a_0+a_-^\dag a_-$.
Although $L^2$ seems SU(2) symmetric, it is not
because $L_x$, $L_y$, and $L_z$ are not genuine
angular moment operators (for spin-1 atoms); they do not
satisfy the Casimir relation $L^2\ne N(N+1)$ \cite{ozgur}.
As was shown before \cite{pu,ozgur} multi-atom
internal state correlations
continue to arise dynamically with $H'$ and
the addition of single atom Raman couplings of
the type $i\Omega_{\mu\nu}(a_\mu^\dag a_\nu-a_\nu^\dag a_\mu)/2$.
Unfortunately, we have not been able to
solve for the combined dynamics analytically even
for a small number of atoms. It is also not apparent
how to numerically investigate strategies for
creating a N-GHZ state in this case.
\begin{figure}
\includegraphics[width=2.75in]{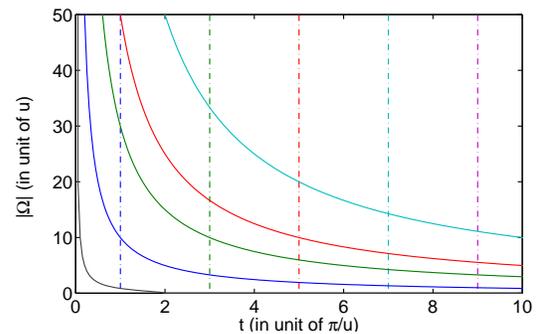}\\
\caption{Solutions of $t_m$ and $\Omega$
as given by the cross
points of the two families of curves $(ut_m)=(2k+1)\pi$
(for $k=0,1,2,3,4$)
and $\sqrt{1+4(\Omega/u)^2}(ut_m)=2m\pi$
(for $m=1,10,30,50,100$). }
\label{fig0}
\end{figure}

Looking back on the two mode model (\ref{h1}) discussed
earlier, we realize that, with a constant
$\Omega$, a state with two atoms
initially in $|b\rangle$ develops into
a 2-GHZ state within a time of $\approx\pi/|u|$.
Specifically, we find
\begin{eqnarray}
C_{11}(t) &&={\Omega\over \sqrt{2} \tilde\Omega}e^{i{u\over 2}t}
\sin\tilde\Omega t, \nonumber\\
C_{20}(t) &&={1\over 2}-i{u\over 4\tilde\Omega}e^{i{u\over 2}t}
\sin\tilde\Omega t
+{1\over 2}  e^{i{u\over 2}t} \cos\tilde\Omega t,\nonumber\\
C_{02}(t) &&={1\over 2}+i {u\over 4\tilde\Omega}e^{i{u\over 2}t}
\sin\tilde\Omega t
-{1\over 2}  e^{i{u\over 2}t} \cos\tilde\Omega t,
\label{sol}
\end{eqnarray}
with $\tilde\Omega=\sqrt{u^2+4\Omega^2}/2$
for the coefficients of state vector expansion
\begin{eqnarray}
|\psi(t)\rangle =&&C_{20}(t){1\over \sqrt 2}b^{\dag 2}|0,0\rangle
+C_{02}(t){1\over \sqrt 2} a^{\dag 2}|0,0\rangle\nonumber\\
&&+C_{11}(t) b^\dag a^\dag |0,0\rangle.
\label{ghzs}
\end{eqnarray}
In the above Eq. (\ref{sol}), we have omitted a
common phase factor $e^{-iut}$. Clearly, $C_{11}(t)=0$ occurs at
\begin{eqnarray}
2\tilde\Omega t_m=\sqrt{1+4(\Omega/u)^2}(ut_m)=2m\pi.
\end{eqnarray}
$|\psi(t)\rangle$ becomes a 2-GHZ state (\ref{ghz})
when $|C_{20}|=|C_{02}|=1/\sqrt{2}$. This occurs at
$ut_m=(2k+1)\pi$ since $C_{20/02}(t_m)=
[1\pm e^{i{u}t_m/2} (-1)^m]/2$.
When $|\Omega/u|\gg 1$ both conditions can be
satisfied at different values of $t_m$ and $\Omega$
as shown in Fig. \ref{fig0}.
The shortest time for a 2-GHZ is then $\sim\pi/|u|$.

Based on this observation, we explored numerically the dynamics
of the Hamiltonian
$H=u(L^2-2N)+i\Omega_{\mu\nu}(a_\mu^\dag a_\nu-a_\nu^\dag a_\mu)/2$
assuming a constant $\Omega_{\mu\nu}$ and all atoms initially
in the $|+\rangle$ state.
As expected, we discovered that maximally entangled states
continue to be generated at times $\sim \pi/|u|$
for $N=2,3,4$.

For $N=2$, we find that we get a 2-GHZ state
$(a_+^{\dag 2}+e^{i\phi}a_{\mu=0,-}^{\dag 2})|0,0,0\rangle/2$
with either a Raman drive $\Omega_{+0}$
or $\Omega_{+-}$. $\phi$ a controllable phase shift.
The 2-GHZ state occurs at times of
$\approx (2k+1)\pi/|u|$ ($\mu=0$) or $(2k+1)\pi/4|u|$
($\mu=-$) and also times shifted
by a small multiples of $\pi/|\Omega_{+\mu}|$
(when $|\Omega_{+\mu}|\gg |u|$) in
their immediate neighborhoods.
The state fidelities are always very high
as long as $k$ is not too large.

For $N=3$, only the $\Omega_{+-}$ drive
seems to create a 3-GHZ state
$\propto (a_+^{\dag 3}+e^{i\phi}a_{-}^{\dag 3})|0,0,0\rangle$
at times differing from $\approx (2k+1)\pi/4|u|$ by small
multiples of $\pi/|\Omega_{+-}|$.
Maximum correlated atomic ensembles in states $|+\rangle$
and $|-\rangle$ were previously
predicted to occur due to elastic
collisions for a initial condensate in
state $|0\rangle$ \cite{epr}.

\begin{figure}
\includegraphics[width=3.25in]{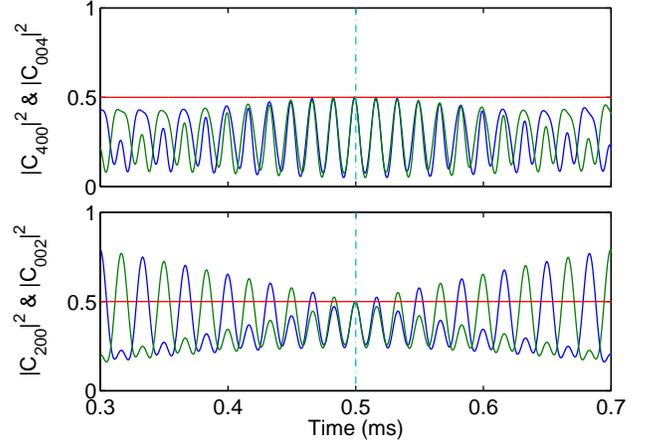}\\
\caption{The two oscillating lines are respectively
the probabilities for all atoms in state
 $|+\rangle$ or $|-\rangle$. Top panel is for $N=4$,
 while the bottom one is for $N=2$ atoms. The vertical dot-dashed
 line is at $t=0.5$ (ms).}
\label{fig}
\end{figure}

For $N=4$, we find that again only the
$\Omega_{+-}$ drive allows for a simple
identification of a 4-GHZ state
$\propto (a_+^{\dag 4}+e^{i\phi}a_{-}^{\dag 4})|0,0,0\rangle$,
which also occurs at $\approx (2k+1)\pi/4|u|$
and values shifted by a small
multiples of $\pi/|\Omega_{+-}|$ in its neighborhood.
Thus at $t\approx \pi/4|u|$ atoms in wells with $N=2$ and $4$
are both maximum entangled as illustrated in
Fig. \ref{fig}.
In this simulation, we have used
$\Omega_{+-}=(2\pi) 30$ (kHz) and $u=(2\pi)0.25$ (kHz).
We note that their respective values
are not important except that they scale inversely
with the required time. What seems to be important is
to assure that $|\Omega_{+-}/u|\ge 100$ for up to
4-atoms to achieve a high fidelity maximum entangled state.

In conclusion, we have presented a simple and efficient
protocol for turning a Mott insulator condensate of
$^{87}$Rb atoms in the ground state $F=1$ manifold into
a source for maximally entangled atomic clusters.
Our protocol is reliable and accessible with
current technologies \cite{hansch}. It
produces maximum entangled quantum states
of Bose-condensed atoms with high fidelity.
The only noticeable drawback seems
to be due to the fact that for $^{87}$Rb atoms,
$u\propto (a_2-a_0)$, i.e. the difference
of scattering lengths for the two symmetric channels
with total spin $0$ and $2$.
Nevertheless, inelastic decay processes are essentially
negligible because
all spin states of the atomic ground state manifold
are included. Furthermore, the N-GHZ state
$(|+\rangle^{\otimes N}+e^{i\phi}|-\rangle^{\otimes N})/\sqrt{2}$
is stable against elastic collisions, which are required
to conserve the total $M_F$, i.e.
atoms in $|+\rangle$ (or $|-\rangle$) state remain in
the same state after collisions.
 Thus the slow dynamics is perhaps
not a major course of concern.
Other atomic
species (e.g. $F=1$ manifold of $^{23}$Na \cite{kurn})
may provider large values of $u$.
In Ref. \cite{qcz}, a quantum logic operation
between two atoms (one each in two neighboring wells)
was proposed that
uses the much stronger (by two orders of magnitudes)
interaction $\propto a_{ab}$. Application
of this in a Mott state (with one atom per well)
produces GHZ states on a faster time scale,
although it requires more complicated
internal state dependent optical trapping.

Finally, a condensate in a Mott state
contains many individual wells with identical number
of atoms \cite{hansch}. This makes the experimental
detection of the entanglement (for atoms within each well)
relatively easy. One can perform the usual parity-type
measurement with Ramsey's oscillatory fields
technique \cite{monroe} (again) by driving the single
atom Raman transition so quickly that collision effects
are negligible.
All wells with the same number of atoms thus
contribute to the detected signal.
Generalizations of our protocol
to more than 4-atoms and other related results
will be published elsewhere.

This work is supported by a grant from NSA, ARDA, and DARPA
under ARO Contract No. DAAD19-01-1-0667, and by a grant
from the NSF PHY-0113831.


\begin{references}
\bibitem{note}
For odd values of $N$, an additional single atom
interaction of the form
$\Omega J_x$ (for a duration of $\pi/2$)
will be sufficient to create a N-GHZ (see Ref. \cite{mol}).
Such a Raman $\pi$ pulse can be implemented
by taking $\Omega\to i\Omega$ in the
$\Omega J_y$ term of Eq. (\ref{h1}).
\end{references}
\end{document}